\documentclass[9pt,twocolumn,twoside]{pnas-new}

\usepackage{authblk}
\usepackage{url}
\usepackage{siunitx}
\usepackage{graphicx}
\usepackage{multirow, framed}
\usepackage{booktabs,caption}
\usepackage{amsmath}
\usepackage{amssymb}
\usepackage{lineno}

\templatetype{pnasresearcharticle}

\title{Robust machine learning segmentation for large-scale analysis of heterogeneous clinical brain MRI datasets}

\author[a,1]{Benjamin Billot}
\author[b]{Colin Magdamo} 
\author[b]{You Cheng} 
\author[b]{Steven E. Arnold}
\author[b]{Sudeshna Das}
\author[a,c,d]{Juan Eugenio Iglesias}

\affil[a]{Centre for Medical Image Computing, University College London, London, UK}
\affil[b]{Department of Neurology, Massachusetts General Hospital and Harvard Medical School, Boston, USA}
\affil[c]{Martinos Center for Biomedical Imaging, Massachusetts General Hospital and Harvard Medical School, Cambridge, USA}
\affil[d]{Computer Science and Artificial Intelligence Laboratory, Massachusetts Institute of Technology, Cambridge, USA}

\leadauthor{Billot} 

\significancestatement{
\textit{SynthSeg$^{+}$} is the first image segmentation tool for automated analysis of highly heterogeneous brain MRI clinical scans. Our method relies on a new strategy to train deep neural networks, such that it can robustly analyse scans of any contrast and resolution without retraining, which was previously impossible. Moreover, \textit{SynthSeg$^{+}$} enables scalable quality control of the produced results by automatic detection of faulty segmentations. Our tool is publicly available with FreeSurfer and can be used ``out-of-the-box'', which facilitates its use and enhances reproducibility. By unlocking the analysis of heterogeneous clinical data, \textit{SynthSeg$^{+}$} has the potential to transform neuroimaging studies, given the considerable abundance of clinical scans compared to the size of datasets~used~in~research.}

\authorcontributions{B.B and J.E.I. took part in the conceptualisation and implementation of this work, the writing of the manuscript, and the software release. Y.C performed visual control of the produced segmentations and edited the manuscript. C.M., S.D., and S.E.A. curated the data and edited the manuscript.}
\authordeclaration{The authors declare no competing interests.}
\correspondingauthor{\textsuperscript{1}E-mail: benjamin.billot.18\@ucl.ac.uk}


\begin{abstract}
Every year, millions of brain MRI scans are acquired in hospitals, which is a figure considerably larger than the size of any research dataset. Therefore, the ability to analyse such scans could transform neuroimaging research. Yet, their potential remains untapped, since no automated algorithm is robust enough to cope with the high variability in clinical acquisitions (MR contrasts, resolutions, orientations, artefacts, subject populations). Here we present \textit{SynthSeg$^{+}$}, an AI segmentation suite that enables, for the first time, robust analysis of heterogeneous clinical datasets. In addition to whole-brain segmentation, \textit{SynthSeg$^{+}$} also performs cortical parcellation, intracranial volume estimation, and automated detection of faulty segmentations (mainly caused by scans of very low quality). We demonstrate \textit{SynthSeg$^{+}$} in seven experiments, including an ageing study on 14,000 scans, where it accurately replicates atrophy patterns observed on data of much higher quality. \textit{SynthSeg$^{+}$} is publicly released as a ready-to-use tool to unlock the potential of quantitative morphometry.
\end{abstract}

\dates{This manuscript was compiled on \today}
\doi{\url{www.pnas.org/cgi/doi/10.1073/pnas.XXXXXXXXXX}}

\begin{document}

\maketitle
\thispagestyle{firststyle}
\ifthenelse{\boolean{shortarticle}}{\ifthenelse{\boolean{singlecolumn}}{\abscontentformatted}{\abscontent}}{}

\section*{Introduction}

\paragraph{Background.}

Neuroimaging plays a prominent role in our attempt to understand the human brain, as it enables an array of analyses such as volumetry, morphology, connectivity, physiology, and molecular studies. A prerequisite for almost all these analyses is the contouring of brain structures, a task known as image segmentation. In this context, magnetic resonance imaging (MRI) is the imaging technique of choice, since it enables the acquisition of noninvasive scans \textit{in vivo} with excellent soft-tissue contrast.

The vast majority of neuroimaging studies rely on prospective datasets of high-quality MRI scans, and especially on \SI{1}{\milli\meter} T1-weighted acquisitions. Indeed, these scans present a remarkable white-grey matter contrast, and can be easily analysed with widespread neuroimaging packages (SPM~\cite{ashburner_unified_2005} FSL~\cite{jenkinson_2012_fsl} FreeSurfer~\cite{fischl_freesurfer_2012}) to derive quantitative morphometric measurements. Meanwhile, brain MRI scans acquired in the clinic (e.g., for diagnostic purposes) present much higher variability in acquisition protocols, and thus cannot be analysed with conventional neuroimaging softwares. This variability is threefold. First, clinical scans use a wide range of MR sequences and contrasts, which are chosen depending on the tissue properties to highlight. Then, they often present real-life artefacts that are uncommon in research datasets, such as very low signal-to-noise ratio, or incomplete field of view. Finally, instead of using 3D scans at high resolution like in research, physicians usually prefer to acquire a sparse set of 2D images in parallel planes, which are faster to inspect but introduce considerable variability in terms of slice spacing, thickness, and orientation.

The ability to analyse clinical datasets is highly desirable, since they represent the overwhelming majority of brain MRI scans. For example, 10 million brain clinical scans were acquired in the US in 2019 alone~\cite{oren_curbing_2019}. This figure is orders of magnitude larger than the size of the biggest research studies such as ENIGMA~\cite{hibar_common_2015} or UK BioBank~\cite{alfaro-almagro_image_2018}, which comprise tens of thousands subjects. Hence, analysing such clinical data would considerably increase the sample size and statistical power of the current neuroimaging studies. Furthermore, it would also enable the analysis of populations that are currently underrepresented in research studies (e.g., UK BioBank and ADNI~\cite{jack_alzheimers_2008} with 95\% white subjects~\cite{adni_sdni_2012,fry_comparison_2017}), but that are more easily found in clinical datasets. Therefore, there is a clear need for an automated segmentation tool that is robust to MR contrast, resolution, clinical artefacts, and subject populations.

\paragraph{Related Works.}

The gold standard in brain MRI segmentation is manual delineation. However, this tedious procedure requires costly expertise and is untenable for large-scale clinical applications. Alternatively, one could only consider high-quality scans (i.e., \SI{1}{\milli\meter} T1-weighted scans) that can be analysed with neuroimaging softwares, but this would drastically decrease effective sample sizes, because such scans are expensive and seldom acquired in the clinic.

Several methods have been proposed for segmentation of MRI scans of variable contrast or resolution. First, contrast-adaptiveness has classically been addressed with Bayesian strategies using unsupervised likelihood model~\cite{puonti_fast_2016}. Nevertheless, the accuracy of these methods progressively deteriorates at decreasing resolutions due to partial volume effects, where voxel intensities become less representative of the underlying tissues~\cite{choi_partial_1991}. While such effects can theoretically be modelled within the Bayesian framework~\cite{van_leemput_unifying_2003}, the resulting algorithm quickly becomes intractable at decreasing resolutions, thus precluding analysis of clinical scans with large slice thickness.

The modern segmentation literature mostly relies on supervised convolutional neural networks (CNNs)~\cite{milletari_v-net_2016,ronneberger_u-net_2015}, which obtain fast and accurate results on their training domain (i.e., scans with similar contrast and resolution). However, CNNs suffer from the ``domain-gap'' problem~\cite{pan_survey_2010}, where networks do not generalise well to data with different resolution~\cite{ghafoorian_transfer_2017} or MR contrast~\cite{akkus_deep_2017}, even within the same modality (e.g., T1-weighted scans acquired with different parameters or hardware)~\cite{karani_test-time_2021}. Data augmentation techniques have addressed this problem in intra-modality scenarios by applying spatial and intensity transforms to the training data~\cite{zhang_generalizing_2020}. However, the resulting CNNs still need to be retrained for each new MR contrast or resolution, which necessitates costly labelled images. Another approach to bridging the domain gap is domain adaptation, where CNNs are explicitly trained to generalise from a ``source'' domain with labelled data, to a specific ``target'' domain, where no labelled examples are available~\cite{chen_synergistic_2019,karani_test-time_2021}. Although these methods alleviate the need for supervision in the target domain, they still need to be retrained for each new domain, which makes them impractical to apply at scale on highly heterogeneous clinical data.

Very recently, we proposed \textit{SynthSeg}~\cite{billot_synthseg_2021}, the first method that can segment brain scans of any contrast and resolution without retraining. This was achieved by adopting a domain randomisation approach~\cite{tobin_domain_2017}, where a 3D CNN is trained on synthetic scans of fully randomised contrast and resolution. Consequently, \textit{SynthSeg} learns domain-agnostic representations, which provide it with an outstanding generalisation ability compared with previous methods~\cite{billot_synthseg_2021}. However, \textit{SynthSeg} frequently falters when applied to clinical scans with low signal-to-noise ratio, poor tissue contrast, or acquired at very low resolution -- an issue that we address in the present article.

Several strategies have been introduced to improve the robustness of CNNs, most notably hierarchical models. These models divide the final task into easier operations such as: progressive refining of segmentations at increasing resolutions~\cite{isensee_nnu-net_2021}, or segmenting the same image with increasingly finer labels~\cite{roth_application_2018}. Although hierarchical models can help improving performance, they may still struggle to produce topologically plausible segmentations in difficult cases, which is a well-known problem for CNNs~\cite{nosrati_incorporating_2016}. Recent approaches have sought to solve this problem by modelling high-order topological relations, either by aligning predictions and ground truths in latent space during training~\cite{oktay_anatomically_2018}, by correcting predictions with a registered atlas~\cite{sinclair_2022_atlas}, or with denoising networks~\cite{larrazabal_post-dae_2020}.

While the aforementioned methods substantially improve robustness, they do not guarantee accurate results in every case. Hence, the ability to identify erroneous predictions is crucial, especially when analysing clinical data of varying quality. Traditionally, this has been achieved with visual quality control (QC), but several automated strategies have now been proposed to replace this tedious procedure. A first class of methods seek to register predictions to a pool of reference segmentations to compute similarity scores~\cite{valindria_reverse_2017}, but the required registrations remain time-consuming. Therefore, recent techniques employ fast CNNs to model QC as a regression task, where faulty segmentations are rejected by applying a thresholding criterion on the regressed scores~\cite{kohlberger_evaluating_2019,liu_alarm_2019,wang_deep_2020}.

\paragraph{Contributions.}

In this article, we present \textit{SynthSeg$^{+}$}, the first clinical brain MRI segmentation suite that is robust to MR contrast, resolution, clinical artefacts and to a wide range of subject populations. Specifically, the proposed method leverages a novel deep learning architecture composed of hierarchical networks and denoisers. This new architecture is trained on synthetic data with the domain randomisation approach introduced by \textit{SynthSeg}, and is shown to considerably increase the robustness of the original method to clinical artefacts. Furthermore, \textit{SynthSeg$^{+}$} includes new modules for cortex parcellation, automated failure detection, and estimation of intracranial volume (ICV, a crucial covariate in volumetry). All aspects of our method are thoroughly evaluated on more than 15,000 highly heterogeneous clinical scans, where \textit{SynthSeg$^{+}$} is shown to enable, for the first time, automated segmentation and volumetry of large, uncurated clinical datasets. A first version of this work was presented at MICCAI 2022, for whole-brain segmentation only~\cite{billot_robust_2022}. Here we considerably extend our previous conference paper by adding cortical parcellation, automated QC, and ICV estimation, as well as by evaluating our approach in five new experiments. The proposed method can be run with FreeSurfer using the simple following command:

\begin{verbatim}
mri_synthseg --i [input] --o [output] --robust
\end{verbatim}

% -------------------------------------------------------------
% -------------------------- Results --------------------------
% -------------------------------------------------------------

\section*{Results}

% ------ overview ------

\begin{figure*}[t]
\centering
\includegraphics[width=0.9\textwidth]{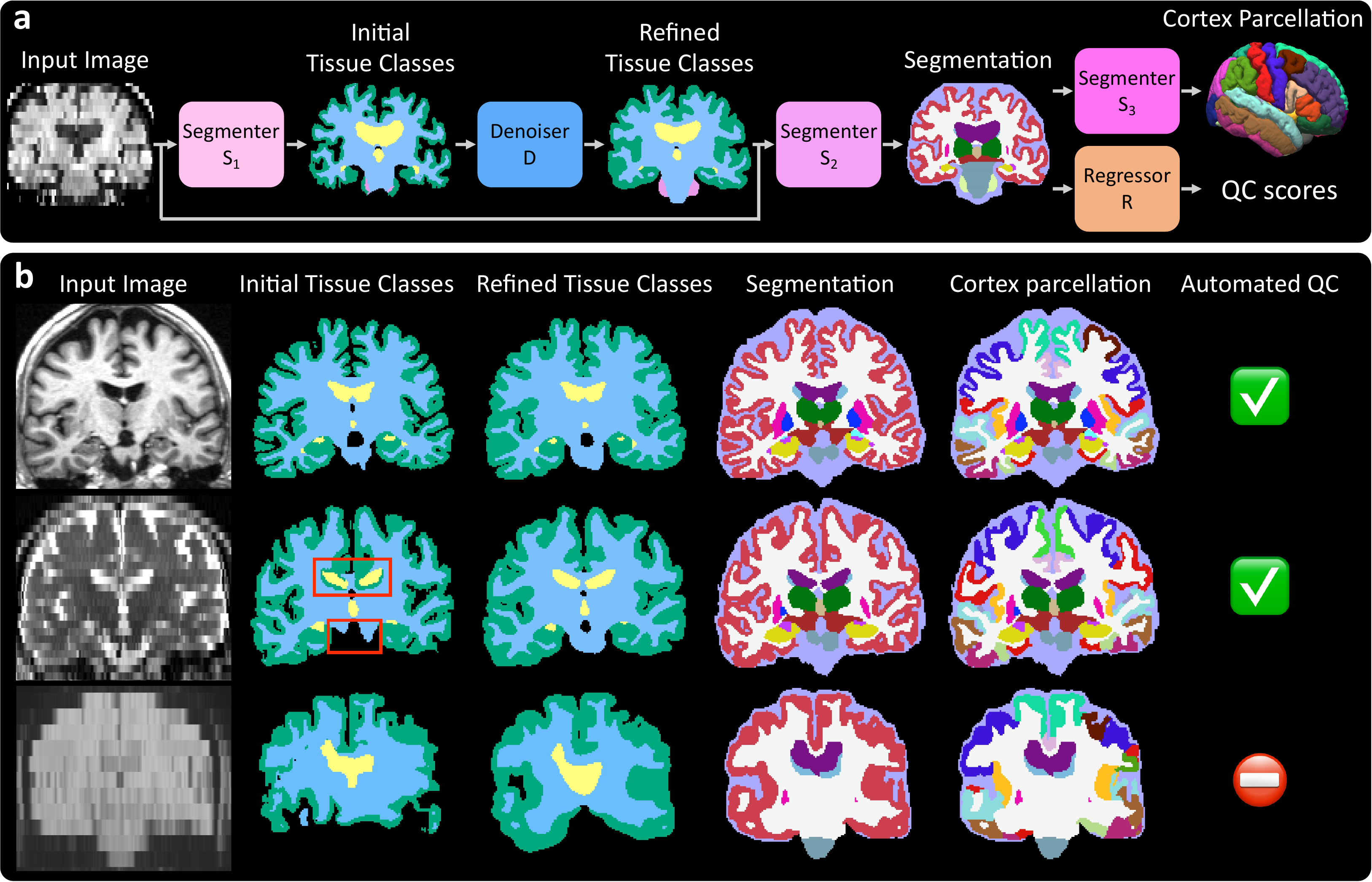}
\caption{Overview of \textit{SynthSeg$^{+}$}. (a) Inference pipeline. All modules are implemented as CNNs. (b) Outputs of the intermediate modules for three representative cases. On the first row, all modules obtain accurate results. On the second row, the denoiser corrects mistakes in the initial tissue classes (red boxes), ultimately leading to a good segmentation. Third, the very low tissue contrast of the image leads to a poor segmentation, but the automated QC correctly identifies it as unusable for subsequent analyses.}
\label{fig:overview}
\end{figure*}

\paragraph{A multi-task segmentation suite.}
\textit{SynthSeg$^{+}$} segments uni-modal brain MRI scans by using hierarchical modules designed to efficiently decompose the segmentation task into easier intermediate steps that are less prone to errors (Figure~\ref{fig:overview}). Specifically, a first network $S_1$ produces preliminary segmentations of four coarse tissue classes, which are corrected for potential topological mistakes by a denoiser $D$ (see red boxes in Figure~\ref{fig:overview}.b). Next, predictions of the target labels are obtained with a segmenter $S_2$. The obtained cortical region is then further parcellated using a segmenter $S_3$. Finally, a regressor $R$ provides us with ``QC scores'' (describing the quality of the obtained segmentations) for 10 regions, which we use to perform automated QC. We emphasise that all segmentations are given at \SI{1}{\milli\meter} isotropic resolution, regardless of the native resolution of the input.

In the following experiments, we first quantitatively evaluate the accuracy of \textit{SynthSeg$^{+}$} for whole-brain segmentation and cortex parcellation. Next, we assess the performance of the automated QC module for automatic failure detection. We then present results obtained for ICV estimation, and a proof-of-concept volumetric study for detection of Alzheimer's Disease patients. Finally, we demonstrate \textit{SynthSeg$^{+}$} in a real-life study of ageing conducted on more than 14,000 uncurated and heterogeneous clinical scans.

% ------ whole-brain, clinical scans ------

\begin{figure*}[t]
\centering
\includegraphics[width=0.9\textwidth]{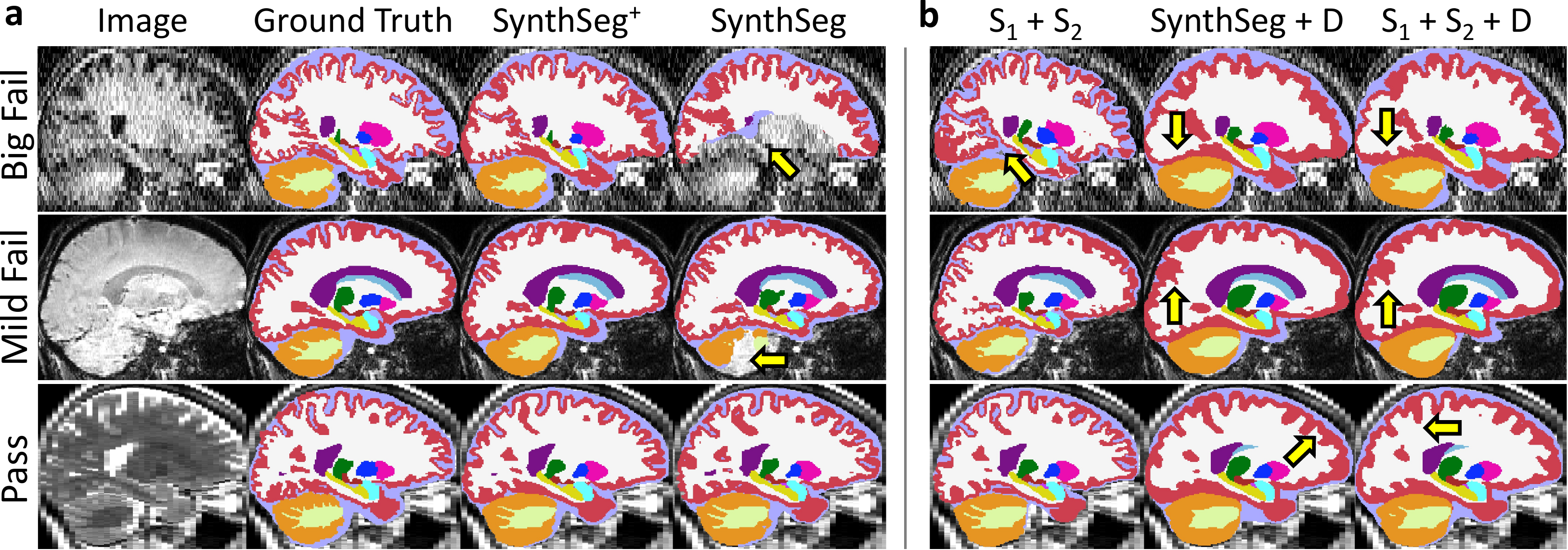}
\caption{Segmentations obtained by all tested methods. (a) Comparison between whole-brain segmentations produced by \textit{SynthSeg$^+$} and \textit{SynthSeg}. Here we show the results obtained for three cases, where \textit{SynthSeg} respectively exhibits large (``big fail''),  moderate (``mild fail''), and no errors (``pass''). Yellow arrows point at notable mistakes. \textit{SynthSeg$^{+}$} produces excellent results given the low SNR, poor tissue contrast, or low resolution of the input scans. (b) Segmentations obtained on the same scans by three variants of our method. Note that appending $D$ substantially smooths segmentations.}
\label{fig:segmentations}
\end{figure*}

\begin{figure*}[t]
\centering
\includegraphics[width=\textwidth]{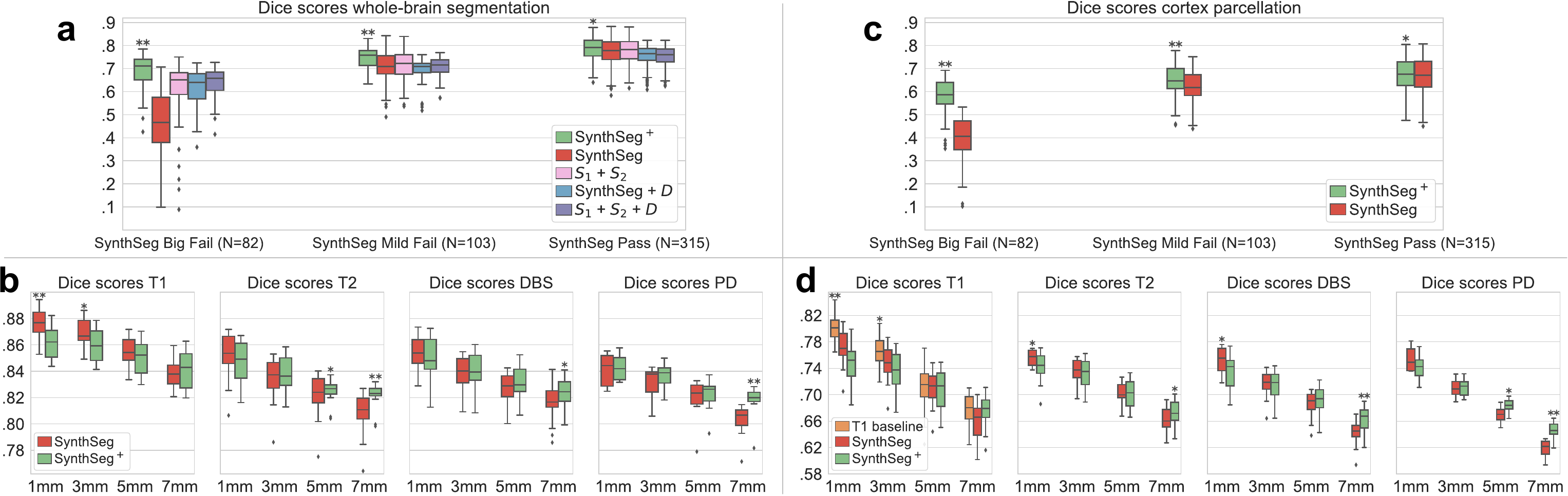}
\caption{Dice scores for whole-brain segmentation (a,b) and cortical parcellation (c,d). For (a) and (c), we evaluate the competing methods on 500 heterogeneous clinical scans, presented based on a visual QC of \textit{SynthSeg} segmentations. The results in (b) and (d) are obtained on scans of four MRI modalities at decreasing resolutions. For each dataset, the best method is marked with $\ast$ or $\ast\ast$ if statistically better than the others at a 5\% or 1\% level (one-sided Bonferroni-corrected Wilcoxon signed-rank test).}
\label{fig:dice}
\end{figure*}

\paragraph{Whole-brain segmentation on clinical data.} 
In this first experiment, we quantitatively assess the accuracy of \textit{SynthSeg$^{+}$} for whole-brain segmentation of clinical acquisitions. In this purpose, we use 500 heterogeneous labelled scans, that we take from the picture archiving communication system (PACS) of Massachusetts General Hospital. We compare \textit{SynthSeg$^{+}$} to \textit{SynthSeg}~\cite{billot_synthseg_2021} and three ablations. First, we evaluate an architecture representative of classical cascaded networks ($S_1 + S_2$), by ablating the denoiser $D$. Then, we test two more variants obtained by appending a denoiser to \textit{SynthSeg} (\textit{SynthSeg$+ D$}) and the cascaded networks ($S_1 + S_2 + D$), with the aim of evaluating state-of-the-art postprocessing denoisers~\cite{larrazabal_post-dae_2020}. We measure accuracy with Dice scores, which quantify the overlap between predicted and reference segmentations.

For visualisation purposes, we split the results into three classes based on a visual QC performed on the segmentations of \textit{SynthSeg}: ``big fails'', ``mild fails'' and ``passes'' (Figure~\ref{fig:segmentations}). The results, shown in Figure~\ref{fig:dice}a, reveal that the hierarchical design of \textit{SynthSeg$^{+}$} considerably improves robustness (with a mean 76 Dice points), and yields the best scores in all three categories. \textit{SynthSeg$^{+}$} shows an outstanding improvement of 23.5 Dice points over \textit{SynthSeg} for big fails, and outperforms it by 5.1 and 2.4 points on mild fails and passes, respectively.

In comparison with cascaded networks, employing a denoiser $D$ in \textit{SynthSeg$^{+}$} to correct the mistakes of $S_1$ consistently improves the results by 2.3 to 5.1 Dice points. Additionally, using $D$ within our framework (rather than for postprocessing) enables us to substantially outperform the two other variants by 3.1 to 6.2 Dice points. This outcome is explained by the fact that denoisers return very smooth segmentations, especially for the convoluted cortex (Figure~\ref{fig:segmentations}). While the denoiser $D$ of our method also exhibits important smoothing effects (Figure~\ref{fig:overview}), we emphasise that these are successfully recovered by $S_2$, which produces segmentations with sharp boundaries.

% ------ whole-brain, resolution study ------

\paragraph{Robustness against MR contrast and resolution.}
We now evaluate the performance of \textit{SynthSeg$^{+}$} as a function of MR contrast and resolution. Here, we use high-resolution research scans (\SI{1}{\milli\meter} isotropic resolution) of four MR contrasts, and artificially downsample them at progressively decreasing resolutions. Specifically, we downsample 20 T1-weighted, 18 T2-weighted, 8 proton density (PD), and 18 deep brain stimulation (DBS) scans to nine different resolutions: 3, 5 and \SI{7}{\milli\meter} in axial, coronal, and sagittal orientation.

The results are displayed in Figure~\ref{fig:dice}b and show that both \textit{SynthSeg} and \textit{SynthSeg$^{+}$} maintain a high level of accuracy across all tested contrasts and resolutions (above 80 Dice points). The two methods yield very similar scores, but with different trends: while \textit{SynthSeg} produces slightly sharper and more accurate segmentations at \SI{1}{\milli\meter} isotropic resolution (up to 1.4 Dice points better), \textit{SynthSeg$^{+}$} remains more robust at lower resolutions, where it obtains superior scores for all MRI modalities (maximum gap of 1.6 Dice points).

% ------ cortex parc. ------

\paragraph{Cortex parcellation.} 
Here, we assess the accuracy of \textit{SynthSeg$^{+}$} for cortex parcellation, and compare it against the results obtained by appending the cortical parcellation module $S_3$ to \textit{SynthSeg}. Figure~\ref{fig:dice}c shows that, as in the first experiment, \textit{SynthSeg$^{+}$} vastly improves the results of \textit{SynthSeg} on heterogeneous clinical data: it yields better scores by 17.1 Dice points on big fails, and remains superior by 1.8 points on good cases. Regarding the evolution of performance at decreasing resolutions, Figure~\ref{fig:dice}d confirms the trend observed for whole-brain segmentation: although \textit{SynthSeg} is more accurate at high resolution (largest gap of 1.6 Dice points for \SI{1}{\milli\meter} T1-weighted scans), it is outperformed by \textit{SynthSeg$^{+}$} at lower resolutions (2.6 points for \SI{7}{\milli\meter} proton density scans). We note that the scores for cortex parcellation are below those obtained for whole-brain segmentation, which is not surprising since cortical regions are smaller and more convoluted, and thus more difficult to segment. Nevertheless, the quality of the present results is remarkable, since they are similar to the scores obtained by a state-of-the-art supervised CNN trained on T1-weighted scans~\cite{zhang_generalizing_2020}.

% ------ QC ------

\begin{table}[t]
\centering
\setlength\tabcolsep{3pt} 
    \caption{Results of the automated QC analysis on 594 clinical heterogeneous scans for \textit{SynthSeg$^{+}$} and two competitors. The best score for each metric is in bold. No statistical difference is found for the AUC between Liu et al.~\cite{liu_alarm_2019} and our regressor R (p=0.081 when computing a DeLong test~\cite{delong_comparing_1988}).}
    \begin{tabular}{l c c c c c}
        \toprule
         Method & Sensitivity & Specificity & Accuracy & AUC \\
        \midrule
        $D$ vs. $S_2$ outputs & 0.936 & 0.668 & 0.884 & 0.858 \\
        Liu \textit{et al.}~\cite{liu_alarm_2019} & 0.982 & 0.906 & 0.968 & 0.989 \\
        Regressor $R$ (\textit{SynthSeg$^{+}$}) & \textbf{0.999} & 0.906 & \textbf{0.981} & \textbf{0.998} \\
        \bottomrule
        \label{tab:qc}
    \end{tabular}
\end{table}

\paragraph{Automated quality control.}
We now test the automated QC module of \textit{SynthSeg$^{+}$} on the 500 clinical scans employed in the previous experiments plus 94 new scans that were almost unusable due to insufficient field of view, wrong organs, critical artefacts, etc. These 594 scans are segmented with \textit{SynthSeg$^{+}$}, and visually classified by an expert neuroanatomist (Y.C.) on a pass/fail basis. This analysis results in a rejection rate of 18.2\% for \textit{SynthSeg$^{+}$}, while this rate falls to a remarkable 4.9\% when excluding the unusable scans (see examples of fails in Supplementary 2). We now seek to automatically replicate the results of this visual QC. Here, \textit{SynthSeg$^{+}$} is set to reject a segmentation if at least one region obtains an automated QC score below 0.65. We compare \textit{SynthSeg$^{+}$} against two competitors: a state-of-the-art technique for regression of QC scores~\cite{liu_alarm_2019}, and a simpler version of this work, where segmentation quality is estimated by computing Dice scores between the outputs of $D$ and $S_2$ (the labels of the latter being converted to the four tissue classes).

Table~\ref{tab:qc} reports the sensitivity, specificity, accuracy, and area under the ROC curve (AUC)~\cite{bradley_use_1997} obtained by each method for this binary classification task (ROC curves are given in the Supplementary materials). Despite its relative simplicity, the approach comparing the outputs of $D$ and $S_2$ already yields a fair accuracy: it correctly classifies 88.4\% of the cases, albeit with limited specificity (66.8\%). While the two other strategies vastly improve the results (with accuracies above 96\%), the simple regressor used in \textit{SynthSeg$^{+}$} obtains scores very similar to the dedicated state-of-the-art method~\cite{liu_alarm_2019}: although no statistical difference can be inferred between AUCs (DeLong test~\cite{delong_comparing_1988}), our approach slightly outperforms Liu et al. for all metrics other than specificity.

% ------ ICVs ------

\begin{figure}[t]
\centering
\includegraphics[width=0.9\columnwidth]{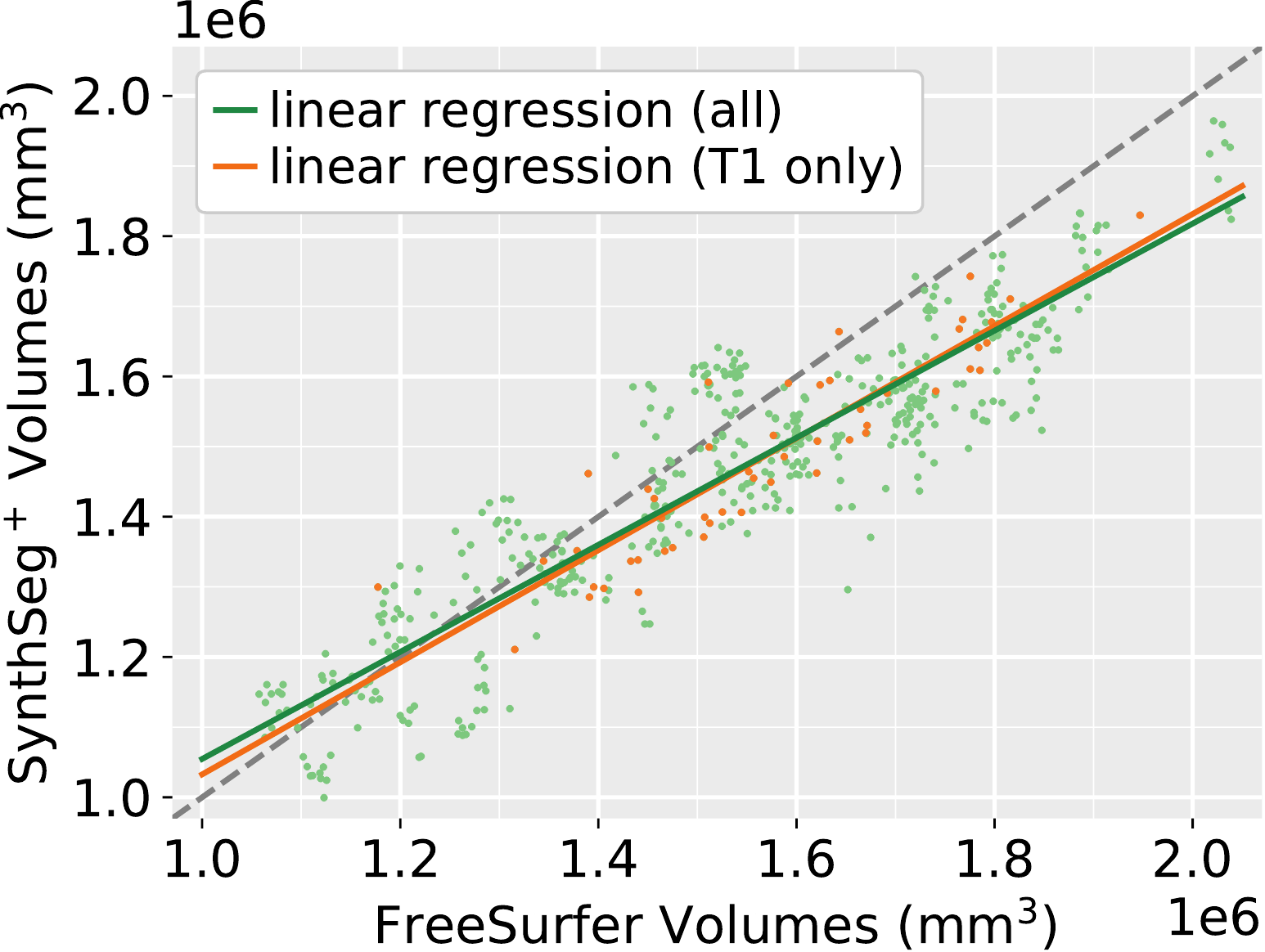}
\caption{Scatter plot of ICVs predicted by \textit{SynthSeg$^{+}$} and FreeSurfer~\cite{fischl_freesurfer_2012} on the 500 clinical scans. Orange points depict 1mm T1-weighted acquisitions (N=62), while the other scans are in green. The grey dashed line marks where abscissa is equal to ordinate. The Pearson correlation coefficient between the two methods is 0.910 when considering all scans, and 0.906 for T1-weighted scans only (p<10\textsuperscript{-9} in both cases).}
\label{fig:icvs}
\end{figure}

\paragraph{Intracranial volume estimation.} 
ICV estimation is a crucial task in volumetric studies, since it is used to correct head-size effects. Here, we use the clinical dataset of 500 scans to compute correlation coefficients between the ICVs estimated with \textit{SynthSeg$^{+}$} and FreeSurfer~\cite{fischl_freesurfer_2012}. First, we focus on a subset of 62 T1-weighted scans at \SI{1}{\milli\meter} resolution, since analysing these high-quality scans will provide us with a theoretical upper bound for correlation. For these scans, Figure~\ref{fig:icvs} illustrates that both methods produce strongly correlated ICVs (Pearson's $r=0.910$). Remarkably, results hardly change when extending this analysis to all the 500 scans ($r=0.906$, $p = 0.071$ for a two-sided signed-rank Wilcoxon test between the two ICV distributions), which highlights the robustness of our ICV estimation module against clinical data of varying quality. Finally, a closer inspection of Figure~\ref{fig:icvs} reveals that \textit{SynthSeg$^{+}$} tends to predict lower values than FreeSurfer for bigger heads. This observation is consistent with the literature, where similar results have been reported when comparing FreeSurfer to ICVs derived from manual segmentations~\cite{malone_2015_accurate}.

% ------ AD vol. study ------

\paragraph{Alzheimer's Disease volumetric study.} 
We now conduct a proof-of-concept volumetric analysis to study whether \textit{SynthSeg$^{+}$} can detect subtle hippocampal atrophy linked with Alzheimer's Disease~\cite{gosche_hippocampal_2002}. Here, we use a new dataset of 100 subjects, half of whom are diagnosed with Alzheimer's Disease. All subjects are imaged with \SI{1}{\milli\meter} T1-weighted scans, and fluid-attenuated inversion recovery (FLAIR) scans at \SI{5}{\milli\meter} axial resolution, which enables us to study performance across images of varying quality. The predicted hippocampal volumes are linearly corrected for age, gender, and ICV (estimated with \textit{SynthSeg$^{+}$}). Differences between control and diseased populations are then measured by computing effect sizes (Cohen's d~\cite{cohen_statistical_1988}).

Table~\ref{tab:cohens_d} reports the results obtained by FreeSurfer, \textit{SynthSeg} and \textit{SynthSeg$^{+}$}. It shows that all methods yield strong effect sizes (between 1.36 and 1.40) on the T1-weighted scans. While segmenting the hippocampus is of modest complexity in \SI{1}{\milli\meter} T1-weighted scans, this task becomes much more challenging on the axial FLAIR scans, where the hippocampus only appears in very few slices (often just 2 to 4). Nonetheless, both \textit{SynthSeg} and \textit{SynthSeg$^{+}$} maintain a remarkable level of accuracy, and are still able to detect strong effect sizes (1.23 and 1.20, respectively). We emphasise that the results obtained are very similar across methods, which is confirmed by the absence of statistical difference when running DeLong tests on corresponding AUCs (all p-values are above 0.3).

\begin{table}[!t]
\centering
\setlength\tabcolsep{7pt} 
    \caption{Effect sizes for hippocampal volumes predicted by FreeSurfer, \textit{SynthSeg}, and \textit{SynthSeg$^{+}$} between 50 controls and 50 Alzheimer's Disease patients. All subjects were imaged with 1mm T1-weighted scans, as well as 5mm axial FLAIR scans. AUC scores obtained by every method are shown in parentheses. All approaches produce very similar results, and no statistical difference can be inferred (DeLong tests on AUCs all result in p-values above 0.3).}
    \begin{tabular}{l c c c c c}
        \toprule
         Contrast & Resolution & FreeSurfer & \textit{SynthSeg} & \textit{SynthSeg$^{+}$}  \\
        \midrule
        T1 & 1mm$^3$ & 1.38 (0.895) & 1.40 (0.898) & 1.36 (0.891) \\
        FLAIR & 5mm axial & - & 1.23 (0.876) & 1.20 (0.872) \\
        \bottomrule
    \label{tab:cohens_d}
    \end{tabular}
\end{table}

% ------ ageing study ------

\begin{figure*}[!ht]
\centering
\includegraphics[width=0.85\textwidth]{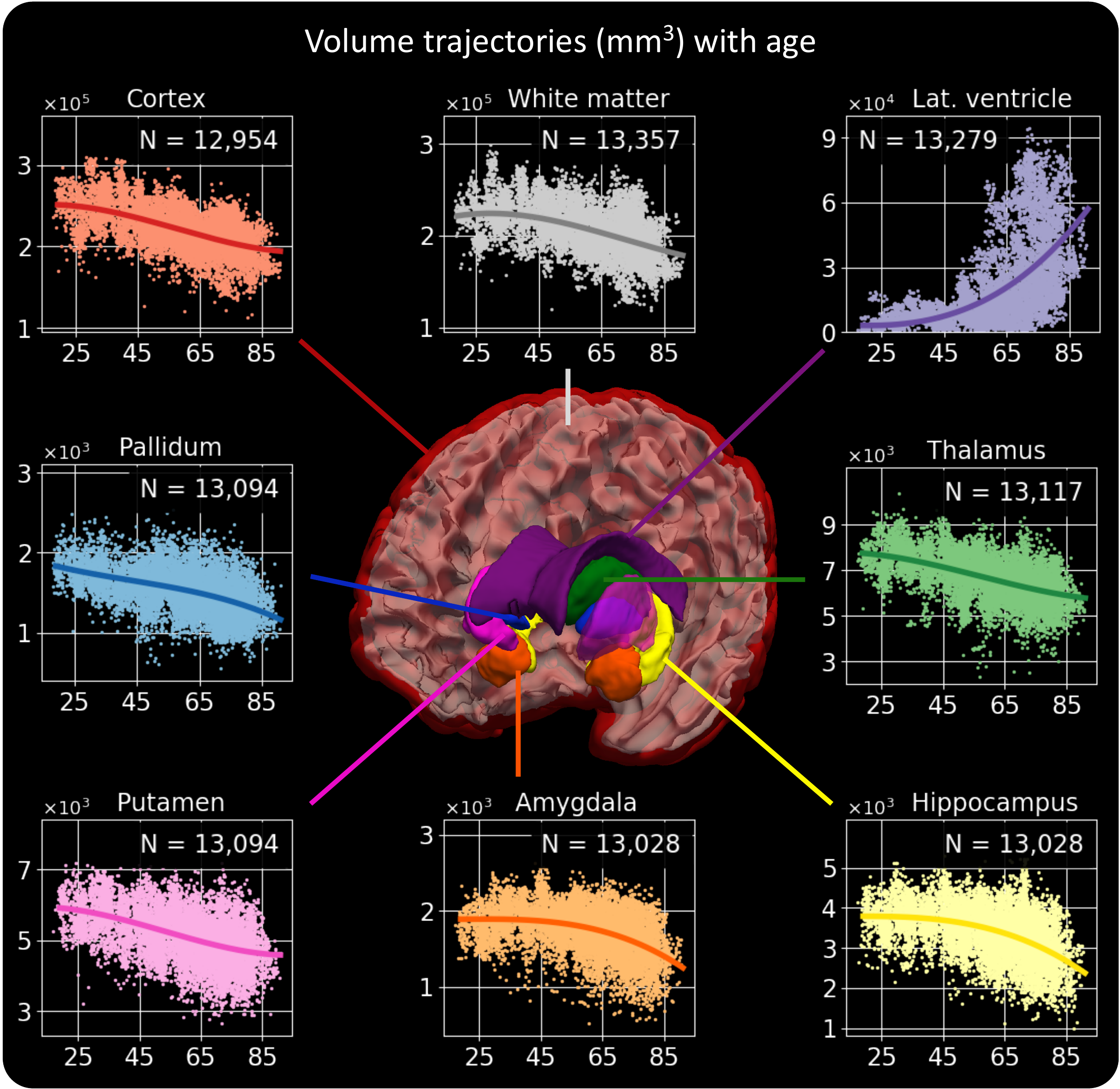}
\caption{Volume trajectories obtained by processing 14,752 heterogeneous clinical scans with SynthSeg$^+$. For each brain region, the value of $N$ indicates the number of volumes considered to build the plot, i.e., the number of segmentations that passed the automated QC for this structure. We emphasise that the obtained results are remarkably similar to recent studies, which exclusively employed scans of much higher quality~\cite{bethlehem_brain_2022,coupe_towards_2017,dima_subcortical_2022}.}
\label{fig:ageing_all}
\end{figure*}

\begin{figure*}[!ht]
\centering
\includegraphics[width=\textwidth]{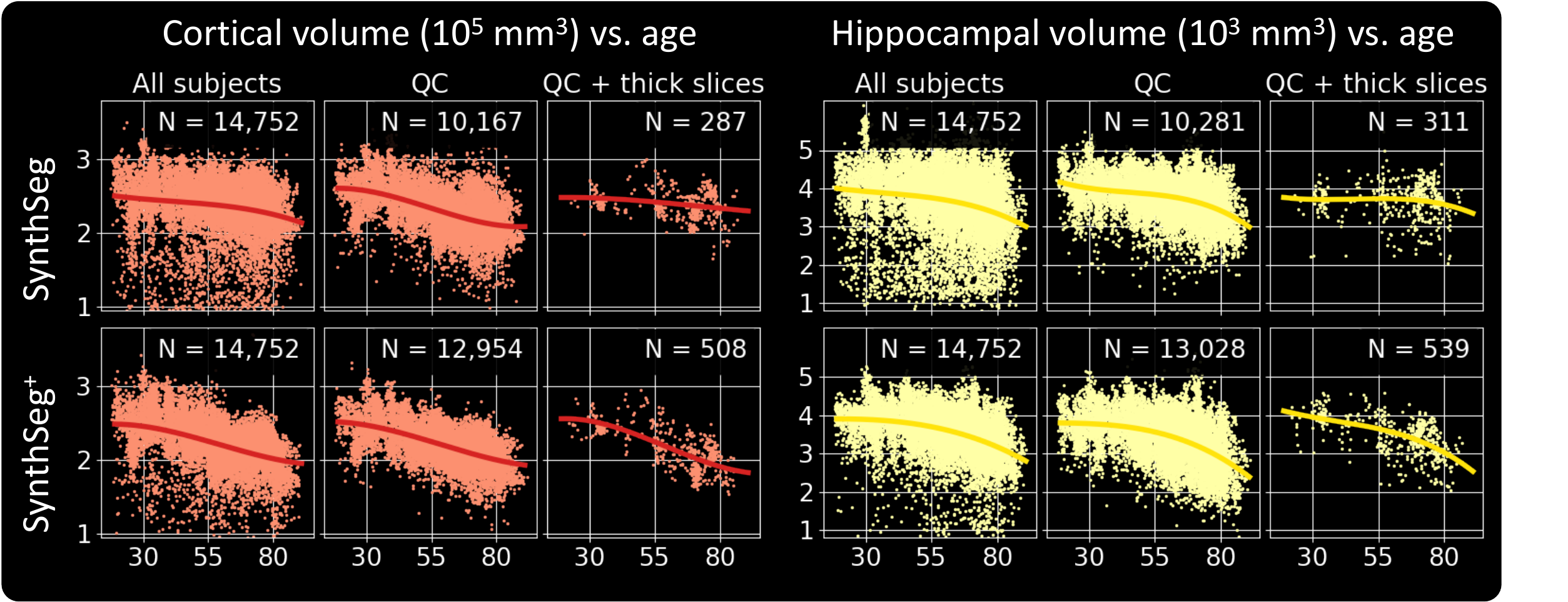}
\caption{Cortical and hippocampal volume trajectories obtained by \textit{SynthSeg} and \textit{SynthSeg$^+$}. This includes different scenarios: using all available scans, keeping only those which passed the automated QC (N = 13,028), and simulating the case where we only have access to scans acquired at low resolution (here with a slice thickness of more than 6.5mm). We highlight that \textit{SynthSeg$^+$} is much more robust than \textit{SynthSeg} since it outputs far fewer outliers, and is able to accurately detect atrophy patterns even for scans at very low resolutions.}
\label{fig:ageing_cortex}
\end{figure*}

\paragraph{Ageing study on over 14,000 clinical scans.}
In this final experiment, we conduct a proof-of-concept clinical volumetric study on 14,752 scans from 1,367 patients with neurology visits at the Massachusetts General Hospital (see Materials and Supplementary 1 for detailed information). Specifically, we verify whether \textit{SynthSeg$^{+}$} is able to reproduce well-known ageing atrophy patterns. We first process all scans with \textit{SynthSeg$^{+}$}, and filter the results with the automated QC module. Instead of rejecting whole segmentations based on a global threshold, we now apply a per-structure rejection criterion, where we only keep the volumes corresponding to regions with a QC score above 0.65. This provides us with between 12,954 (cortex) and 13,357 (white matter) volumes for each region. We then build age-volume trajectories independently for each region with a B-spline model, with linear correction for gender and scan resolution. Remarkably, Figure~\ref{fig:ageing_all} shows that the obtained trajectories are highly similar to the results obtained by recent studies on scans of much higher quality (i.e., \SI{1}{\milli\meter} T1-weighted scans)~\cite{bethlehem_brain_2022,coupe_towards_2017,dima_subcortical_2022}. For example, \textit{SynthSeg$^{+}$} accurately replicates: the peak in white-matter volume at approximately 30 years-old, the acute increase in ventricular volume for ageing subjects, and the early onset of thalamic atrophy compared to the hippocampus and amygdala. 

If we now compare it to \textit{SynthSeg}, the proposed approach exhibits similar trajectories, but produces far fewer outliers (Figure~\ref{fig:ageing_cortex}). This can be seen by the substantially cleaner curves obtained by \textit{SynthSeg$^{+}$} when including volumes from all available scans, or by the considerably higher number of segmentations that passed the automated QC. Finally, as opposed to \textit{SynthSeg}, \textit{SynthSeg$^{+}$} produces almost identical results when we  evaluate it exclusively on scans of very low-resolution (i.e., with slice spacing above \SI{6.5}{\milli\meter}), which further highlights its robustness against scans of widely varying quality.

% -------------------------------------------------------------
% ------------------------ Discussion -------------------------
% -------------------------------------------------------------

\section*{Discussion}

Here we present \textit{SynthSeg$^{+}$}, the fist segmentation suite for large-scale analysis of highly heterogeneous clinical brain MRI scans. The proposed method leverages a novel deep learning architecture, where state-of-the-art domain-agnostic CNNs perform the target segmentation task in a hierarchical fashion. As a result, \textit{SynthSeg$^{+}$} exhibits an unparalleled robustness to clinical artefacts, and is the first method that can accurately segment clinical data of any MR contrast and resolution, including scans with poor tissue contrast and low signal-to-noise ratio. Moreover, the two proof-of-concept volumetric studies and the quantitative ageing experiment (Supplementary 4) have demonstrated the robustness of the proposed method against a wide range of subject populations. Finally, our highly precise QC enables us to automatically discard the few erroneous segmentations from subsequent analyses.

The performance of our hierarchical modules is demonstrated throughout all experiments, where it considerably improves the robustness of \textit{SynthSeg} to clinical artefacts, both quantitatively (higher Dice scores in the first experiment), and qualitatively (far fewer outliers in the ageing study). Further ablation studies show that using a denoiser $D$ to correct the mistakes of $S_1$ leads to a consistent improvement over cascaded networks. Moreover, \textit{SynthSeg$^{+}$} substantially outperforms state-of-the-art denoising networks for postprocessing~\cite{larrazabal_post-dae_2020}, which is explained by two reasons. First, these methods do not have access to the input scans, and may thus produce segmentations that deviate from the original anatomy. In contrast, we predict final segmentations by exploiting both the input test scans and prior information given by $D$. Second, denoiser predictions are often excessively smooth. Here, we mitigate this issue by using the outputs of $D$ as robust priors for $S_2$, which effectively learns to refine the smooth boundaries given by the denoiser. However, Figure~\ref{fig:segmentations} illustrates that the predictions of \textit{SynthSeg$^{+}$} may still exhibit slight smoothing effects. While the resolution studies show that this residual smoothness leads to a marginally lower accuracy than \textit{SynthSeg} for scans at high resolution (relatively uncommon in clinical settings), we consider this a minor limitation compared to the considerable gain in robustness of the proposed approach.

\textit{SynthSeg$^{+}$} is also the first method for volumetric cortex parcellation of clinical scans in the wild: only FreeSurfer~\cite{fischl_freesurfer_2012} and FastSurfer~\cite{henschel_fastsurfer_2020} could tackle this problem automatically, but they only apply to \SI{1}{\milli\meter} T1-weighted scans. The high accuracy of our parcellation module is demonstrated by evaluating it on high-quality scans, for which it yields competitive scores with state-of-the-art supervised CNNs, either tested on the same scans (i.e., T1-baseline~\cite{zhang_generalizing_2020}), or on different datasets~\cite{henschel_fastsurfer_2020}. Remarkably, \textit{SynthSeg$^{+}$} maintains this high level of performance across all tested domains, including the highly heterogeneous clinical acquisitions. We emphasise that this outstanding generalisation ability can difficultly be matched by supervised CNNs, which would need to be retrained on every single domain using costly training labelled data.

The results have also shown that the proposed regression-based QC strategy accurately detects the few cases where \textit{SynthSeg$^{+}$} fails, which are mainly due to acquisitions of poor quality (e.g., very low SNR, or insufficient coverage of the brain). Interestingly, our method produces slightly better results than the tested state-of-the-art approach~\cite{liu_alarm_2019}, despite the fact that the latter was shown to outperform strategies based on direct regression in a simpler problem (3D segmentation with only one label)~\cite{liu_alarm_2019}. This outcome can be explained by the fact that variational auto-encoders (which are at the core of \cite{liu_alarm_2019}) have been shown in our experiments to produce excessive smoothness when dealing with whole-brain segmentations. We also emphasise that other methods for automated QC were not evaluated in this work, since they rely on lengthy iterative processes~\cite{valindria_reverse_2017,wang_deep_2020}, which makes them impractical to use at the scale of the target clinical datasets.

Finally, we have demonstrated the true potential of \textit{SynthSeg$^{+}$} on more than 14,000 uncurated clinical scans, where it accurately replicates volume trajectories observed on data of much higher quality. This result suggests that \textit{SynthSeg$^{+}$} can be used to investigate other population effects on huge amounts of clinical data, which will considerably increase the statistical power of the current research studies. Moreover, \textit{SynthSeg$^{+}$} also unlocks other potential applications, such as the analysis of scans acquired with the promising portable low-field MRI scanners, or the introduction of quantitative morphometry in the clinic for diagnosing and monitoring diseases.

We emphasise that \textit{SynthSeg$^+$} can be run ``out of the box'' on brain MRI scans of any contrast and resolution, which has three main benefits. First, it greatly facilitates the use of our method, since it eliminates the need for retraining, and thus the associated requirements in terms of labelled data, deep learning expertise, and hardware. Second, relying on a single model considerably improves the reproducibility of the results, since no hyperparameter tuning is required. And finally, it makes \textit{SynthSeg$^+$} easier to disseminate, which is done here by distributing it with the publicly available package FreeSurfer.

Future work will first focus on surface placement to estimate cortical thickness, which is a powerful biomarker for the progression of many neuropsychiatric disorders. While this task may be challenging when analysing low resolution clinical scans (since the slice thickness can be locally larger than cortical folds), we believe that measurements can still be extracted in regions where the surface is orthogonal to the acquisition direction. Then, we will seek to extend \textit{SynthSeg$^+$} to analyse multi-modal data. So far, we have dealt with multi-modal acquisitions by processing all channels separately, but combining them into a common framework could improve accuracy. Finally, even though the ageing clinical study has demonstrated the applicability of \textit{SynthSeg$^+$} to large cohorts with high morphological variability, future work will aim to precisely quantify performance against various pathologies.

Overall, by enabling robust and reproducible analysis of heterogeneous clinical brain MRI scans, we believe that the present work will enable the development of clinical neuroimaging studies with sample sizes considerably higher than those found in research, which has the potential to revolutionise our understanding of the healthy and diseased human brain.

% -------------------------------------------------------------
% -------------------------- Methods --------------------------
% -------------------------------------------------------------

\section*{Materials and Methods}

\paragraph{Training datasets and population robustness.}
The proposed method is trained only on synthetic data (no real images) generated from a set of brain segmentation maps. Here we use 1,020 maps obtained from \SI{1}{\milli\meter} T1-weighted scans: 20 from the OASIS database~\cite{marcus_open_2007}, 500 from the Alzheimer’s Disease Neuroimaging Initiative (ADNI)~\cite{jack_alzheimers_2008}, and 500 from the Human Connectome Project (HCP)~\cite{van_essen_human_2012}. These segmentations contain labels for 31 brain structures, obtained manually (OASIS) or with FreeSurfer (HCP and ADNI)~\cite{fischl_whole_2002}. Moreover, we complement each map with 11 automated labels for extra-cerebral regions~\cite{puonti_accurate_2020}, and 68 FreeSurfer labels for cortex parcellation. We emphasise that, while HCP subjects are young and healthy, ADNI contains ageing and diseased subjects, who frequently exhibit large atrophy and white matter lesions. Thus, using such a diverse population enables us to build robustness across a wide range of morphologies.

\paragraph{Brain MRI test datasets.}
Our experiments feature three datasets. The first one comprises 15,346 clinical scans from the PACS of Massachusetts General Hospital (see detailed information in Supplementary 1). Briefly, these scans are from 1,367 MRI sessions of distinct subjects with memory complaints, between 18 and 90 years old: 749 males (age $= 62.2 \pm 15.2$) and 618 females (age $= 58.1 \pm 17.2$). Importantly, all scans are uncurated, and span a huge range of MR contrasts (T1-weighted, T2-weighted, FLAIR, diffusion MRI, etc.). Acquisitions are isotropic (11\%), and anisotropic in axial (81\%), coronal (4\%), and sagittal (4\%) orientations. The resolution of isotropic scans varies between 0.3 and \SI{4.7}{\milli\meter}. For anisotropic scans, in-plane resolution ranges between 0.2 and \SI{4.7}{\milli\meter}, while slice spacing varies between 0.8 and \SI{10.5}{\milli\meter}. Ground truths for whole-brain segmentation, cortex parcellation, and ICV estimation were obtained for a subset of scans as follows. First, we isolated all sessions ($N=62$) with \SI{1}{\milli\meter} isotropic T1-weighted scans. These were then labelled using FreeSurfer~\cite{fischl_freesurfer_2012}, and the obtained segmentations were rigidly registered~\cite{modat_fast_2010} to the other scans of the corresponding sessions. Note that we also obtained ICVs for all these scans by reporting the estimations given by FreeSurfer on the T1-weighted scans. Finally, we conducted a visual QC on the results, and removed all scans where even a small segmentation error could be seen, either due to FreeSurfer or registration errors (138 scans) and/or poor image quality (94 cases with e.g., insufficient coverage of the brain, wrong organ). In total, this provided us with ground truth segmentations for 520 scans, that we split between validation (20), and testing (500). All the other 14,752 scans were held-out for indirect evaluation.

The second dataset consists of 66 scans from three sub-datasets: 20 T1-weighted scans from OASIS database~\cite{marcus_open_2007}; 18 subjects imaged twice with T2-weighted acquisitions and a sequence typically used in deep brain stimulation (DBS)~\cite{iglesias_probabilistic_2018}; and 8 proton density scans~\cite{fischl_sequence-independent_2004}. All scans are at \SI{1}{\milli\meter} isotropic resolution, and are available with manual or semi-automated segmentations for 31 brain regions~\cite{fischl_whole_2002}. Labels for cortex parcellation were obtained by running FreeSurfer on the T1-weighted scans, or on companion \SI{1}{\milli\meter} T1-weighted acquisitions for the T2-weighted, DBS, and proton density scans.

The last dataset is another subset of 100 scans from ADNI~\cite{jack_alzheimers_2008}, including 47 males and 53 females, aged 72.9 $\pm$ 7.6 years. Half of the subjects are healthy, while the others are diagnosed with Alzheimer's Disease. All subjects are imaged with two acquisitions: \SI{1}{\milli\meter} isotropic T1-weighted scans, and FLAIR scans at \SI{5}{\milli\meter} axial resolution. Volumetric measurements for individual regions and ICVs are retrieved for all subjects by processing the T1-weighted scans with FreeSurfer.

\paragraph{Deep hierarchical segmentation.}
The proposed architecture relies on four CNN modules, which are designed to split the target segmentation task into easier intermediate operations. Specifically, a first segmenter $S_1$ is trained to produce initial segmentations of four coarse labels (cerebral white matter, cerebral grey matter, cerebrospinal fluid, and cerebellum). These labels group brain regions of similar tissue types and intensities, and are thus easier to discriminate than individual structures. The output of $S_1$ is then fed to a denoising network $D$~\cite{larrazabal_post-dae_2020}, which seeks to correct potential topological inconsistencies and larger segmentation mistakes that sometimes occur for scans with poor tissue contrast, or low signal-to-noise ratio. Final whole-brain segmentations are obtained by a second segmenter $S_2$, which takes as inputs the image and the corrected preliminary segmentations. As such, $S_2$ learns to segment the 31 individual target regions by using the coarse tissue segmentations as priors. We note that $S_2$ is given the opportunity to refine the boundaries produced by $D$, which are sometimes excessively smooth. Finally, the segmentations of $S_2$ are completed by passing them to a fourth network $S_3$, which is trained to parcellate the left and right cortex labels from $S_2$ into 68 different substructures.

\paragraph{Training of the segmenters.}
The three segmentation CNNs are trained separately with extremely diverse uni-modal synthetic data created from anatomical segmentation maps by using a generative model inspired by Bayesian segmentation. Crucially, in order to train domain-agnostic networks, we adopt the domain randomisation strategy introduced in our previous work~\cite{billot_synthseg_2021}. Specifically, we draw all the parameters governing the generative model at each minibatch from uniform priors of very large range. Hence, the segmenters are exposed to vastly changing examples in terms of shape, MR contrast, resolution, artefacts (bias field, noise), and even morphology (due to population variability in the training set). As a result, the segmenters are forced to learn contrast and resolution-agnostic features, which enables them to be applied to brain MRI scans of any domain, without requiring any retraining.

Briefly, this generative model only requires a set of \SI{1}{\milli\meter} isotropic label maps as inputs, and synthesises training examples as follows. At each minibatch, we randomly select one of the training maps, and geometrically augment it with a random spatial transform~\cite{zhang_generalizing_2020}. Next, a preliminary image is built by sampling a randomised Gaussian mixture model conditionally on the deformed label map~\cite{billot_learning_2020}. The resulting image is then obtained by consecutively applying a random bias field, noise injection, intensity rescaling between 0 and 1, and a random voxel-wise exponentiation~\cite{zhang_generalizing_2020}. In turn, low-resolution and PV effects are modelled with Gaussian blurring and sub-sampling at random low resolution. Finally, training pairs are obtained by: defining the deformed label map as ground truth, and resampling the low-resolution images back to the \SI{1}{\milli\meter} isotropic grid, such that the downstream segmenters are trained to operate at high resolution~\cite{billot_partial_2020}.

\paragraph{Training of the denoiser.}
State-of-the-art methods for denoising and topological correction rely mainly on supervised CNNs that are trained to recover ground truth segmentations from artificially corrupted versions of the same maps~\cite{karani_test-time_2021,khan_deep_2021,larrazabal_post-dae_2020}. However, the strategies used to corrupt the input segmentations are often handcrafted (random erosion and dilation, swapping of labels, etc.), and thus do not accurately capture the type of errors made by the segmentation method to correct. In order to train $D$ with examples that are representative of $S_1$ errors, we instead degrade real images and feed them to the trained (and frozen) $S_1$. $D$ is then trained to map the outputs of $S_1$ back to their ground truth. During training, images are degraded on the fly with operations similar to the training of segmenters: spatial deformation, bias field, voxel-wise exponentiation, simulation of low resolution, and noise injection. However, the ranges of the parameters controlling these corruptions are made considerably wider than for the training of segmenters, in order to more frequently obtain erroneous segmentations from $S_1$, and thus to enrich the training data.

\paragraph{Automated QC module.}
While the proposed architecture considerably improves robustness, it remains important to detect potential erroneous predictions, especially when segmenting clinical scans of varying quality. Hence, we introduce another module for automated failure detection. More precisely, we train a regressing network $R$ to predict ``performance scores'' for 10 representative regions of interest (white matter, cortex, lateral ventricle, cerebellum, thalamus, hippocampus, amygdala, pallidum, putamen, brainstem), based solely on the segmentations produced by $S_2$. Here, the performance scores aim to reflect Dice scores that would have been obtained if the input scans were available with associated ground truths~\cite{hann_quality_2019}. The segmentation of a region is then classified as failed if it obtains a predicted Dice score lower than 0.65 (a value chosen based on the validation set). In practice, $R$ is trained with the same method as $D$, where we degrade real images, segment them with $S_2$, and feed the obtained segmentations to $R$.

\paragraph{Network architectures.}
All segmenters use the same architecture as \textit{SynthSeg}~\cite{billot_learning_2020}, which is based on a 3D UNet~\cite{ronneberger_u-net_2015}. Briefly, it comprises five levels, each consisting of two convolutions, a batch normalisation, and either a max-pooling (contracting path), or upsampling operation (expanding path). Every convolution employs a 3$\times$3$\times$3 kernel and an Exponential Linear Unit activation~\cite{clevert_fast_2016}, except for the last layer, which uses a softmax. The first layer has 24 features, while this number is doubled after each max-pooling and halved after each upsampling. Finally, following the UNet architecture, we use skip-connections across the contracting and expanding paths.

The denoiser uses a similar, but slightly lighter architecture: it only has one convolution per level, and keeps a constant number of 16 features. Moreover, we suppress the skip connections between the top two levels to find a compromise between UNets, where top-level skip-connections can potentially reintroduce erroneous features at late stages of the network; and auto-encoders, with excessive bottleneck-induced smoothness.

Finally, the regression network follows the same architecture as the encoder of the segmentation CNNs, except that it uses 5$\times$5$\times$5 convolutions as in~\cite{wang_deep_2020}, which greatly improved the results on the validation set. Regression scores are then retrieved by appending two more convolutions of 10 features (one for each QC region) and a global max-pooling.

\paragraph{Learning.}
The three segmenters and the denoiser are trained to minimise the average soft Dice loss~\cite{milletari_v-net_2016}. If $Y_k$ represents the soft prediction for label $k \in {1, ...,K}$, and $T_k$ is its associated ground truth, this loss is given by:
\begin{equation}
\label{eq:soft dice}
    \text{Loss}_\text{seg} = 1 - \frac{1}{K}\sum_{k=1}^{K} \frac{2 \times \sum_{x,y,z} Y_k(x,y,z) T_k(x,y,z)}{\sum_{x,y,z} Y_k(x,y,z)^2 + T_k(x,y,z)^2}.
\end{equation}
The regressor is trained using a sum of square loss function. All networks are trained separately with the Adam optimiser~\cite{kingma_adam_2017} using a learning rate of $10^{-5}$. We train each module until convergence, which approximately takes 300,000 steps for the segmenters (seven days on a Nvidia RTX6000), and 50,000 steps for the regressor (one day on the same GPU). All models are implemented in Keras~\cite{chollet_keras_2015} with a Tensorflow backend~\cite{abadi_tensorflow_2016}.

\paragraph{Inference.}
Test scans are automatically resampled to \SI{1}{\milli\meter} isotropic resolution, and their intensities are normalised between 0 and 1. The trained model then predicts soft probabilistic segmentations for all target labels. Finally, hard segmentations are obtained by applying an argmax operation on these soft predictions. Overall, the whole inference pipeline takes between 12 and 16 seconds per scan on a RTX6000 GPU. We emphasise that when presented with multi-modal data, SynthSeg+ segments each channel separately.

\paragraph{Individual volumes and ICV estimation.}
Volumes of individual brain regions are estimated by summing all the values of the corresponding soft predictions, and multiplying the result by the volume of a voxel (\SI{1}{\milli\meter}$^3$). Note that summing soft probabilities rather than hard segmentations enables us to account for segmentation uncertainties and, to a certain extent, for partial voluming~\cite{van_leemput_unifying_2003}. In turn, ICVs are estimated for every subject by summing the predicted volumes of all structures, including the intracranial cerebro-spinal fluid.

\paragraph{Dice scores.}
Parts of the evaluations use hard Dice scores, which measure the overlap between the same region across two hard segmentations. If $X$ and $Y$ are corresponding structures in two segmentations, their hard Dice score is given by:
\begin{equation}
    Dice(X,Y)  = 2 \times \frac{|X \cap Y|}{|X| + |Y|}
\end{equation}
where $|\cdot|$ represents the cardinality of a set. Therefore, Dice scores vary between 0 (no overlap) and 1 (perfect matching).

\paragraph{Cohen's d.}
In this work, effect sizes in hippocampal volume between control and AD populations are measured with Cohen's d~\cite{cohen_statistical_1988}. If $\mu_{C}$, $s_{C}^{2}$ and $\mu_{AD}$, $s_{AD}^{2}$ designate the sample means and variances of two volume populations of size $n_{C}$ and $n_{AD}$, where $C$ stands for Controls, Cohen's d is given by:
\begin{equation}
d = \frac{|\mu_{C}-\mu_{AD}|}{s}, \quad s = \sqrt{\frac{(n_{C}-1)s_{C}^{2} + (n_{AD}-1)s_{AD}^{2}}{n_{C}+n_{AD}-2}}.
\end{equation}
Cohen's d below 0.2 are considered to be small, whereas values above 0.8 indicate large effect sizes~\cite{cohen_statistical_1988}.

\paragraph{Regression model for ageing.}
Our ageing model includes: B-splines with 10 equally spaced knots for age, linear terms for slice spacing in each acquisition direction (i.e., sagittal, coronal, and axial), and a bias for gender. We then fit this model numerically by minimising the sum of squares of the residuals with the L-BFGS-B method~\cite{byrd_limited_1995}.

\paragraph{Data, Materials and Software availability.}
The code is available at https://github.com/BBillot/SynthSeg, and the trained models for \textit{SynthSeg{$^+$}} and \textit{SynthSeg} are distributed with the publicly available neuroimaging package FreeSurfer~\cite{fischl_freesurfer_2012}. The ADNI~\cite{jack_alzheimers_2008}, HCP~\cite{van_essen_human_2012}, and OASIS~\cite{marcus_open_2007} scans used in this study are all taken from publicly available datasets.

\acknow{This work is supported by the European Research Council (ERC Starting Grant 677697), the EPSRC UCL Centre for Doctoral Training in Medical Imaging (EP/L016478/1), the Department of Health's NIHR-funded Biomedical Research Centre UCLH, Alzheimer’s Research UK (ARUK-IRG2019A-003), and the NIH (1R01AG070988, 1RF1MH12319501).}
\showacknow{}

% Bibliography

% -------------------------------------------------------------
% ------------------------- Supplement ------------------------
% -------------------------------------------------------------

\newpage
\onecolumn

\titlefont{\hspace{-0.7cm}
Robust machine learning segmentation for \\large-scale analysis of heterogeneous clinical brain \\ MRI datasets -- Supplementary materials}

\vspace{1cm}
\Authfont{\hspace{-0.4cm}
Benjamin Billot, Colin Magdamo, You Cheng, Steven E. Arnold, Sudeshna Das,  Juan Eugenio Iglesias}

\renewcommand{\thefigure}{S\arabic{figure}}
\renewcommand{\thetable}{S\arabic{table}}

\vspace{2cm}
\section*{Supplement 1: $\:$ Demographic information and imaging protocols for the large clinical cohort.}
\vspace{0.5cm}
\begin{figure}[h]
    \centering
    \includegraphics[width=0.9\textwidth]{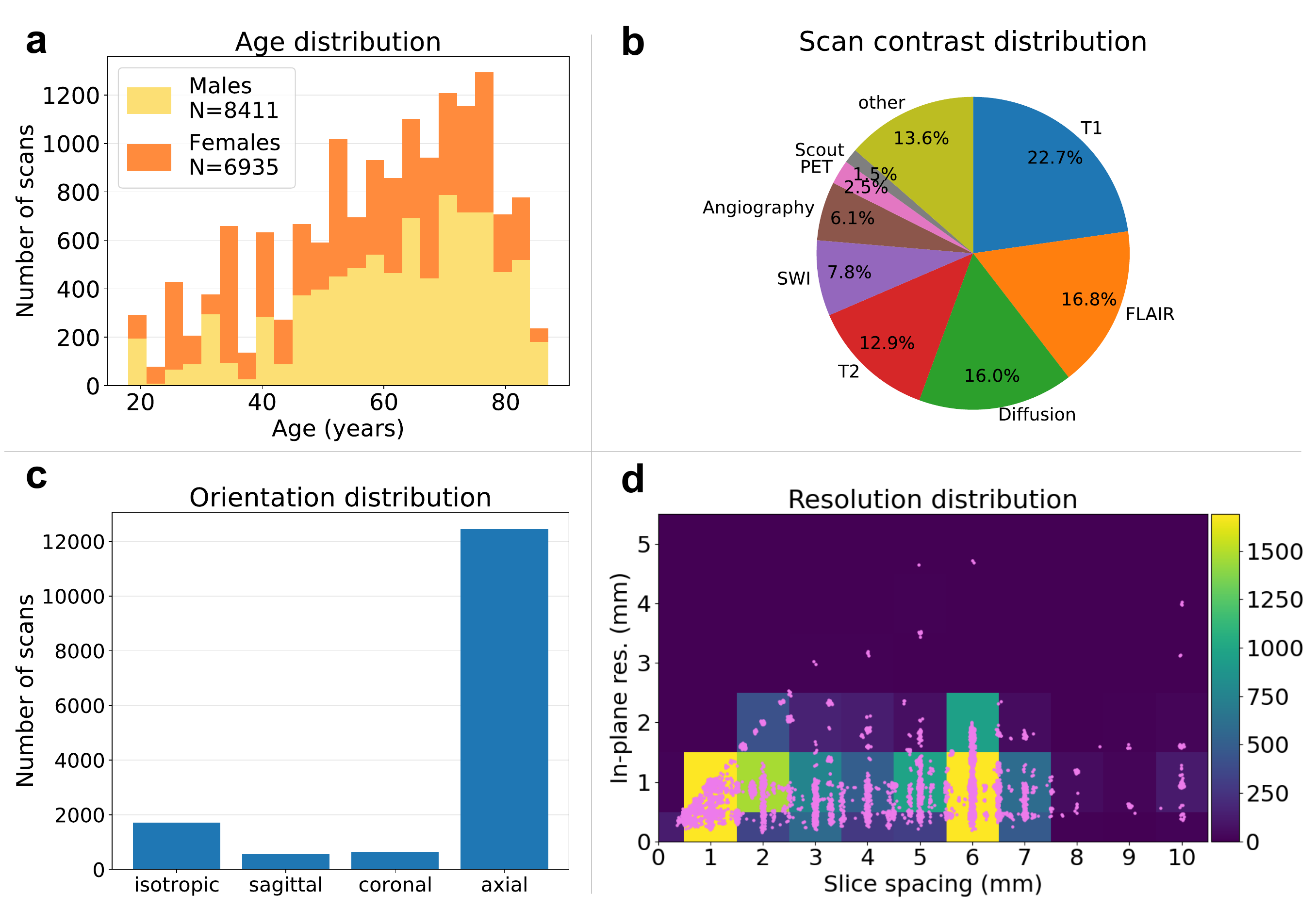}
    \caption{Demographic information and imaging protocols for the 15,346 clinical scans acquired during 1,367 subject sessions at the Massachusetts General Hospital. (a) Age and gender of the subjects, given per scan. (b) Distribution of the used contrast. SWI refers to sensibility weighted imaging, PET to positron emission tomography, and ``other'' to either unidentified contrasts (9.7\%) or sequences with sub-percentage occurrence (e.g., gradient-echo, contrast-agent). (c) Distribution of the acquisition direction (i.e., scan orientation). (d) Distribution of the scan resolution. Individual scans are represented by pink dots over the underlying density distribution.
    }
\end{figure}

\newpage

\section*{Supplement 2: $\:$ Examples of erroneous segmentations by \textit{SynthSeg$^+$}}
\begin{figure}[h]
    \centering
    \includegraphics[width=\textwidth]{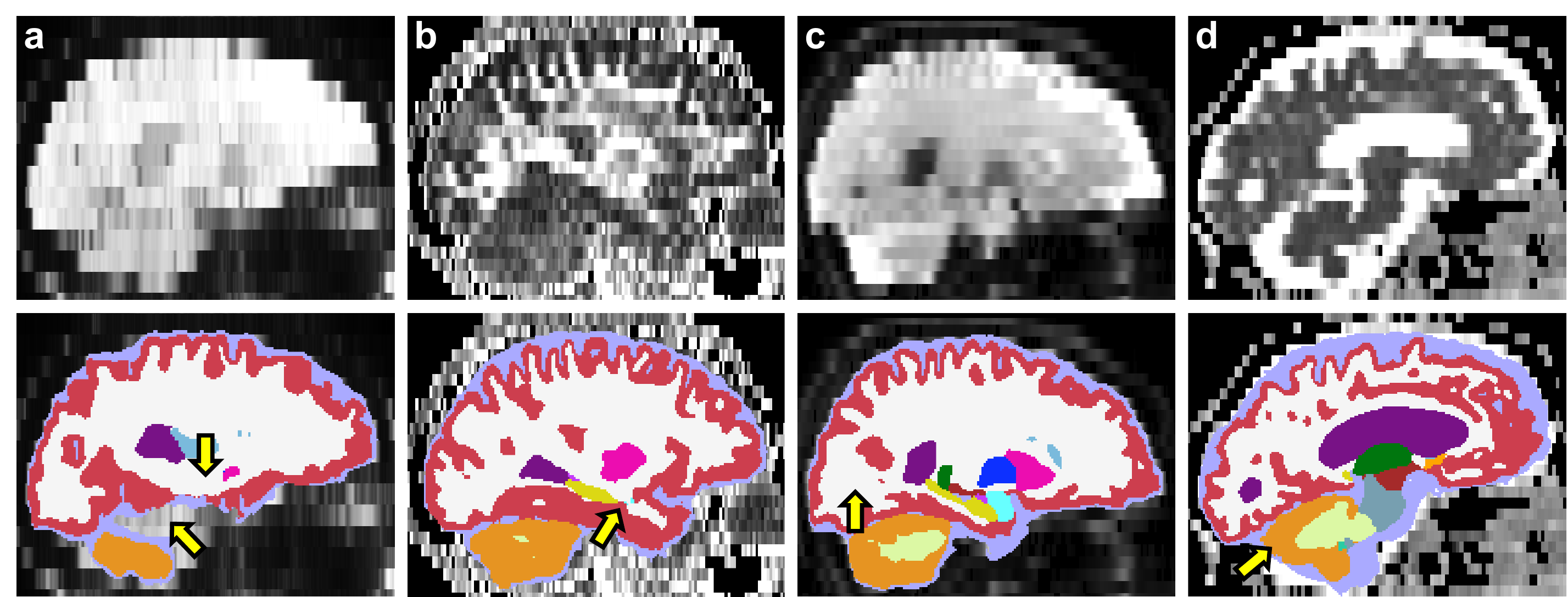}
    \caption{Representative samples of failed segmentations obtained by visual QC. We highlight that these images are very challenging to segment either because of low tissue contrast, low signal-to-noise ratio, or very low resolution. (a) Part of the cerebellum and most subcortical regions are missing. (b) The amygdala is segmented as cerebral cortex (red). (c) Missing posterior lateral ventricle (dark purple), which is instead segmented as white matter (white). (d) The cerebellum is over-segmented in the posterior direction.
    }
\end{figure}

\vspace{2cm}

\section*{Supplement 3: $\:$ ROC curves for the automated QC methods}
\begin{figure}[h]
    \centering
    \includegraphics[width=\textwidth]{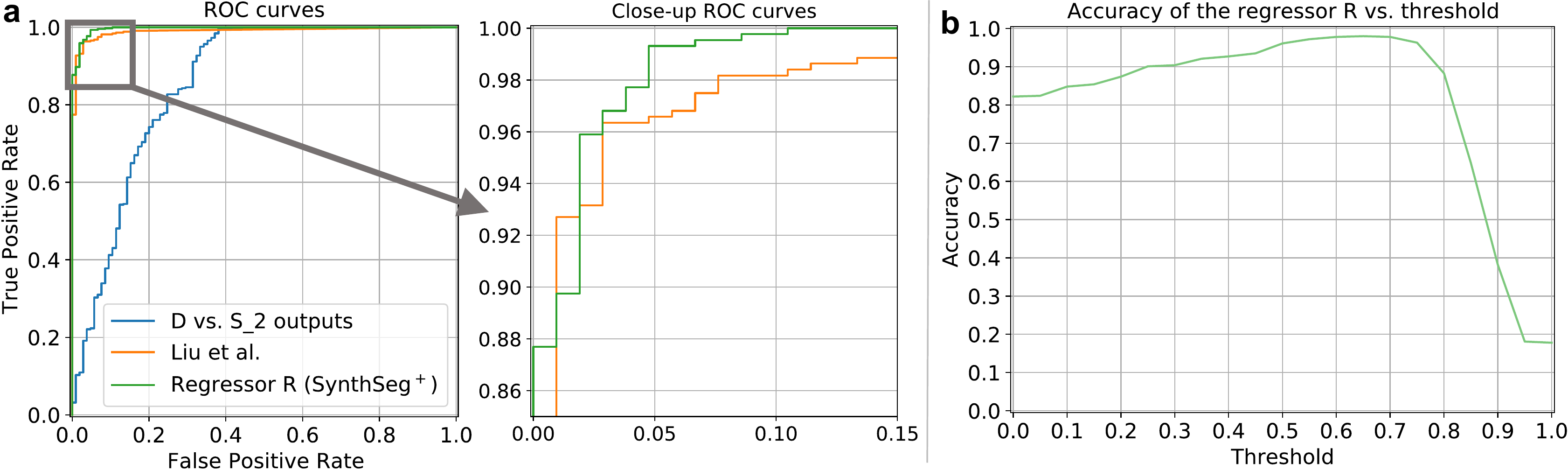}
    \caption{(a) ROC curves of the three methods for automated quality control analysis, with a close-up on the upper left corner. They show that the denoiser-based strategy (i.e., D vs. S\textsubscript{2}) performs the worst. The state-of-the-art method proposed by Liu et al. obtains much better results, but it very slightly outperformed by the regression-based technique implemented in \textit{SynthSeg$^+$}, although no statistical difference is found between the two.(b) Accuracy of our regression-based method as a function of the threshold. Our method is relatively robust to the chosen value, as the accuracy only varies by 0.01 for thresholds between 0.55 and 0.75.
    }
\end{figure}

\newpage

\section*{Supplement 4: $\:$ Quantitative evaluation of the robustness to different subject populations}
\begin{figure}[h]
    \centering
    \includegraphics[width=0.7\textwidth]{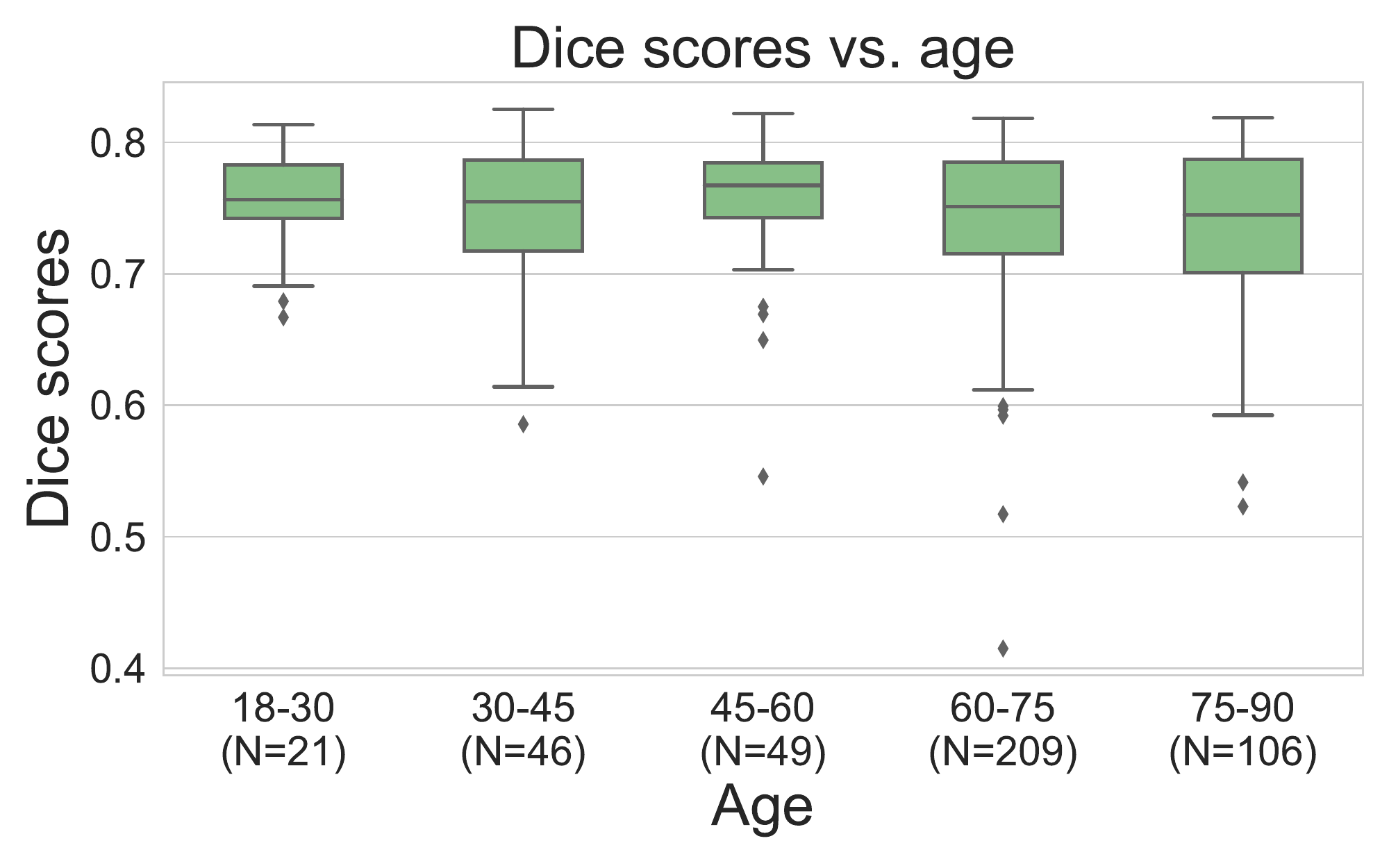}
    \caption{Here we study the accuracy of \textit{SynthSeg$^+$} as a function of age, which we take as a proxy for morphological variability. We plot the Dice scores obtained by the proposed segmentation algorithm on the 500 clinical scans with ground truth. The results show that \textit{SynthSeg$^+$} maintains a high level of accuracy across the whole age range. This is further highlighted by the absence of statistical difference at the p=0.05 level when computing two-sided Wilcoxon signed-rank tests to compare each group with its neighbouring class to the left (minimum p-value of 0.062 between 45-60 and 60-75).
    }
\end{figure}


\begin{thebibliography}{10}

\bibitem{ashburner_unified_2005}
J Ashburner, K Friston, Unified segmentation.
\newblock {\em\protect\JournalTitle{NeuroImage}} \textbf{26}, 39--51 (2005).

\bibitem{jenkinson_2012_fsl}
M Jenkinson, C Beckmann, T Behrens, M Woolrich, S Smith, Fsl.
\newblock {\em\protect\JournalTitle{Neuroimage}} \textbf{62}, 782--790 (2012).

\bibitem{fischl_freesurfer_2012}
B Fischl, {FreeSurfer}.
\newblock {\em\protect\JournalTitle{NeuroImage}} \textbf{62}, 774--781 (2012).

\bibitem{oren_curbing_2019}
O Oren, E Kebebew, J Ioannidis, Curbing {Unnecessary} and {Wasted} {Diagnostic}
  {Imaging}.
\newblock {\em\protect\JournalTitle{JAMA}} \textbf{321}, 245--246 (2019).

\bibitem{hibar_common_2015}
D Hibar, , et~al., Common genetic variants influence human subcortical brain
  structures.
\newblock {\em\protect\JournalTitle{Nature}} \textbf{520}, 224--229 (2015).

\bibitem{alfaro-almagro_image_2018}
F Alfaro-Almagro, , et~al., Image processing and {Quality} {Control} for the
  first 10,000 brain imaging datasets from {UK} {Biobank}.
\newblock {\em\protect\JournalTitle{NeuroImage}} \textbf{166}, 400--424 (2018).

\bibitem{jack_alzheimers_2008}
CR Jack, , et~al., The {Alzheimer}'s {Disease} {Neuroimaging} {Initiative}
  ({ADNI}): {MRI} {Methods}.
\newblock {\em\protect\JournalTitle{Journal of magnetic resonance imaging :
  JMRI}} \textbf{27}, 685--691 (2008).

\bibitem{adni_sdni_2012}
(ADNI), A.s.{D}.{N}.{I}. {Demographic} {Report} (2012).

\bibitem{fry_comparison_2017}
A Fry, , et~al., Comparison of {Sociodemographic} and {Health}-{Related}
  {Characteristics} of {UK} {Biobank} {Participants} {With} {Those} of the
  {General} {Population}.
\newblock {\em\protect\JournalTitle{American Journal of Epidemiology}}
  \textbf{186}, 1026--1034 (2017).

\bibitem{puonti_fast_2016}
O Puonti, JE Iglesias, K Van~Leemput, Fast and sequence-adaptive whole-brain
  segmentation using parametric {Bayesian} modeling.
\newblock {\em\protect\JournalTitle{NeuroImage}} \textbf{143}, 235--249 (2016).

\bibitem{choi_partial_1991}
H Choi, D Haynor, Y Kim, Partial volume tissue classification of multichannel
  magnetic resonance images-a mixel model.
\newblock {\em\protect\JournalTitle{IEEE Transactions on Medical Imaging}}
  \textbf{10}, 395--407 (1991).

\bibitem{van_leemput_unifying_2003}
K Van~Leemput, F Maes, D Vandermeulen, P Suetens, A unifying framework for
  partial volume segmentation of brain {MR} images.
\newblock {\em\protect\JournalTitle{IEEE Transactions on Medical Imaging}}
  \textbf{22} (2003).

\bibitem{milletari_v-net_2016}
F Milletari, N Navab, S Ahmadi, V-{Net}: {Fully} {Convolutional} {Neural}
  {Networks} for {Volumetric} {Medical} {Image} {Segmentation} in {\em
  International {Conference} on {3D} {Vision}}.
\newblock pp. 565--571 (2016).

\bibitem{ronneberger_u-net_2015}
O Ronneberger, P Fischer, T Brox, U-{Net}: {Convolutional} {Networks} for
  {Biomedical} {Image} {Segmentation} in {\em Medical {Image} {Computing} and
  {Computer} {Assisted} {Intervention}}.
\newblock pp. 234--241 (2015).

\bibitem{pan_survey_2010}
S Pan, Q Yang, A {Survey} on {Transfer} {Learning}.
\newblock {\em\protect\JournalTitle{IEEE Transactions on Knowledge and Data
  Engineering}} \textbf{22}, 45--59 (2010).

\bibitem{ghafoorian_transfer_2017}
M Ghafoorian, et~al., Transfer {Learning} for {Domain} {Adaptation} in {MRI}:
  {Application} in {Brain} {Lesion} {Segmentation} in {\em Medical {Image}
  {Computing} and {Computer} {Assisted} {Intervention}}.
\newblock pp. 516--524 (2017).

\bibitem{akkus_deep_2017}
Z Akkus, A Galimzianova, A Hoogi, D Rubin, B Erickson, Deep {Learning} for
  {Brain} {MRI} {Segmentation}: {State} of the {Art} and {Future} {Directions}.
\newblock {\em\protect\JournalTitle{Journal of Digital Imaging}} \textbf{30},
  449--459 (2017).

\bibitem{karani_test-time_2021}
N Karani, E Erdil, K Chaitanya, E Konukoglu, Test-time adaptable neural
  networks for robust medical image segmentation.
\newblock {\em\protect\JournalTitle{Medical Image Analysis}} \textbf{68}
  (2021).

\bibitem{zhang_generalizing_2020}
L Zhang, , et~al., Generalizing {Deep} {Learning} for {Medical} {Image}
  {Segmentation} to {Unseen} {Domains} via {Deep} {Stacked} {Transformation}.
\newblock {\em\protect\JournalTitle{IEEE Transactions on Medical Imaging}}
  \textbf{39}, 2531--2540 (2020).

\bibitem{chen_synergistic_2019}
C Chen, Q Dou, H Chen, J Qin, PA Heng, Synergistic {Image} and {Feature}
  {Adaptation}: {Towards} {Cross}-{Modality} {Domain} {Adaptation} for
  {Medical} {Image} {Segmentation}.
\newblock {\em\protect\JournalTitle{AAAI Conference on Artificial
  Intelligence}} \textbf{33}, 65--72 (2019).

\bibitem{billot_synthseg_2021}
B Billot, , et~al., {SynthSeg}: {Domain} {Randomisation} for {Segmentation} of
  {Brain} {Scans} of any {Contrast} and {Resolution} in {\em arXiv:2107.09559
  [cs]}.
\newblock (2021).

\bibitem{tobin_domain_2017}
J Tobin, et~al., Domain randomization for transferring deep neural networks
  from simulation to the real world in {\em {IEEE}/{RSJ} {International}
  {Conference} on {Intelligent} {Robots} and {Systems}}.
\newblock pp. 23--30 (2017).

\bibitem{isensee_nnu-net_2021}
F Isensee, P Jaeger, S Kohl, J Petersen, K Maier-Hein, {nnU}-{Net}: a
  self-configuring method for deep learning-based biomedical image
  segmentation.
\newblock {\em\protect\JournalTitle{Nature Methods}} \textbf{18}, 203--211
  (2021).

\bibitem{roth_application_2018}
HR Roth, , et~al., An application of cascaded {3D} fully convolutional networks
  for medical image segmentation.
\newblock {\em\protect\JournalTitle{Computerized Medical Imaging and Graphics}}
  \textbf{66}, 90--99 (2018).

\bibitem{nosrati_incorporating_2016}
M Nosrati, G Hamarneh, Incorporating prior knowledge in medical image
  segmentation: a survey in {\em arXiv:1607.01092 [cs]}.
\newblock (2016).

\bibitem{oktay_anatomically_2018}
O Oktay, , et~al., Anatomically {Constrained} {Neural} {Networks} ({ACNNs}):
  {Application} to {Cardiac} {Image} {Enhancement} and {Segmentation}.
\newblock {\em\protect\JournalTitle{IEEE Transactions on Medical Imaging}}
  \textbf{37}, 384--395 (2018).

\bibitem{sinclair_2022_atlas}
M Sinclair, , et~al., Atlas-istn: Joint segmentation, registration and atlas
  construction with image-and-spatial transformer networks.
\newblock {\em\protect\JournalTitle{Medical Image Analysis}} \textbf{78}
  (2022).

\bibitem{larrazabal_post-dae_2020}
A Larrazabal, C Martínez, B Glocker, E Ferrante, Post-{DAE}: {Anatomically}
  {Plausible} {Segmentation} via {Post}-{Processing} {With} {Denoising}
  {Autoencoders}.
\newblock {\em\protect\JournalTitle{IEEE Transactions on Medical Imaging}}
  \textbf{39}, 3813--3820 (2020).

\bibitem{valindria_reverse_2017}
V Valindria, , et~al., Reverse {Classification} {Accuracy}: {Predicting}
  {Segmentation} {Performance} in the {Absence} of {Ground} {Truth}.
\newblock {\em\protect\JournalTitle{IEEE Transactions on Medical Imaging}}
  \textbf{36}, 1597--1606 (2017).

\bibitem{kohlberger_evaluating_2019}
T Kohlberger, V Singh, C Alvino, C Bahlmann, L Grady, Evaluating segmentation
  error without ground truth in {\em Medical {Image} {Computing} and {Computer}
  {Assisted} {Intervention}}.
\newblock pp. 528--536 (2019).

\bibitem{liu_alarm_2019}
F Liu, Y Xia, D Yang, A Yuille, D Xu, An {Alarm} {System} for {Segmentation}
  {Algorithm} {Based} on {Shape} {Model} in {\em {ICCV}}.
\newblock pp. 10652--10661 (2019).

\bibitem{wang_deep_2020}
S Wang, , et~al., Deep {Generative} {Model}-{Based} {Quality} {Control} for
  {Cardiac} {MRI} {Segmentation} in {\em Medical {Image} {Computing} and
  {Computer} {Assisted} {Intervention}}.
\newblock pp. 88--97 (2020).

\bibitem{billot_robust_2022}
B Billot, C Magdamo, SE Arnold, S Das, JE Iglesias, Robust segmentation of
  brain mri in the wild with hierarchical cnns and no retraining in {\em
  Medical {Image} {Computing} and {Computer} {Assisted} {Intervention}}.
\newblock pp. 538--548 (2022).

\bibitem{delong_comparing_1988}
E DeLong, D DeLong, D Clarke-Pearson, Comparing the areas under two or more
  correlated receiver operating characteristic curves: A nonparametric
  approach.
\newblock {\em\protect\JournalTitle{Biometrics}} \textbf{44}, 837--845 (1988).

\bibitem{bradley_use_1997}
AP Bradley, The use of the area under the roc curve in the evaluation of
  machine learning algorithms.
\newblock {\em\protect\JournalTitle{Pattern Recognition}} \textbf{30},
  1145--1159 (1997).

\bibitem{malone_2015_accurate}
I Malone, , et~al., Accurate automatic estimation of total intracranial volume:
  A nuisance variable with less nuisance.
\newblock {\em\protect\JournalTitle{NeuroImage}} \textbf{104}, 366--372 (2015).

\bibitem{gosche_hippocampal_2002}
K Gosche, J Mortimer, C Smith, W Markesbery, D Snowdon, Hippocampal volume as
  an index of alzheimer neuropathology.
\newblock {\em\protect\JournalTitle{Neurology}} \textbf{58}, 1476--1482 (2002).

\bibitem{cohen_statistical_1988}
J Cohen, {\em Statistical {Power} {Analysis} for the {Behavioural} {Sciences}}.
\newblock (Routledge Academic), (1988).

\bibitem{bethlehem_brain_2022}
R Bethlehem, J Seidlitz, S White, J Vogel, K Anderson, Brain charts for the
  human lifespan.
\newblock {\em\protect\JournalTitle{Nature}} \textbf{604}, 525--533 (2022).

\bibitem{coupe_towards_2017}
P Coupé, G Catheline, E Lanuza, J Manjón, Towards a unified analysis of brain
  maturation and aging across the entire lifespan: {A} {MRI} analysis.
\newblock {\em\protect\JournalTitle{Human Brain Mapping}} \textbf{38},
  5501--5518 (2017).

\bibitem{dima_subcortical_2022}
D Dima, , et~al., Subcortical volumes across the lifespan: {Data} from 18,605
  healthy individuals aged 3–90 years.
\newblock {\em\protect\JournalTitle{Human Brain Mapping}} \textbf{43}, 452--469
  (2022).

\bibitem{henschel_fastsurfer_2020}
L Henschel, et~al., Fastsurfer - a fast and accurate deep learning based
  neuroimaging pipeline.
\newblock {\em\protect\JournalTitle{NeuroImage}} \textbf{219} (2020).

\bibitem{marcus_open_2007}
D Marcus, , et~al., {Open Access Series of Imaging Studies (OASIS):
  Cross-sectional MRI Data in Young, Middle Aged, Nondemented, and Demented
  Older Adults}.
\newblock {\em\protect\JournalTitle{Journal of Cognitive Neuroscience}}
  \textbf{19}, 1498--1507 (2007).

\bibitem{van_essen_human_2012}
D Van~Essen, , et~al., The {Human} {Connectome} {Project}: {A} data acquisition
  perspective.
\newblock {\em\protect\JournalTitle{NeuroImage}} \textbf{62}, 2222--2231
  (2012).

\bibitem{fischl_whole_2002}
B Fischl, , et~al., Whole brain segmentation: automated labeling of
  neuroanatomical structures in the human brain.
\newblock {\em\protect\JournalTitle{Neuron}} \textbf{33}, 41--55 (2002).

\bibitem{puonti_accurate_2020}
O Puonti, , et~al., Accurate and robust whole-head segmentation from magnetic
  resonance images for individualized head modeling.
\newblock {\em\protect\JournalTitle{NeuroImage}} \textbf{219} (2020).

\bibitem{modat_fast_2010}
M Modat, , et~al., Fast free-form deformation using graphics processing units.
\newblock {\em\protect\JournalTitle{Computer Methods and Programs in
  Biomedicine}} \textbf{98}, 278--284 (2010).

\bibitem{iglesias_probabilistic_2018}
JE Iglesias, , et~al., A probabilistic atlas of the human thalamic nuclei
  combining ex vivo {MRI} and histology.
\newblock {\em\protect\JournalTitle{NeuroImage}} \textbf{183}, 314--326 (2018).

\bibitem{fischl_sequence-independent_2004}
B Fischl, , et~al., Sequence-independent segmentation of magnetic resonance
  images.
\newblock {\em\protect\JournalTitle{NeuroImage}} \textbf{23}, 69--84 (2004).

\bibitem{billot_learning_2020}
B Billot, et~al., A {Learning} {Strategy} for {Contrast}-agnostic {MRI}
  {Segmentation} in {\em Medical {Imaging} with {Deep} {Learning}}.
\newblock pp. 75--93 (2020).

\bibitem{billot_partial_2020}
B Billot, E Robinson, A Dalca, JE Iglesias, Partial {Volume} {Segmentation} of
  {Brain} {MRI} {Scans} of {Any} {Resolution} and {Contrast} in {\em Medical
  {Image} {Computing} and {Computer} {Assisted} {Intervention}}.
\newblock pp. 177--187 (2020).

\bibitem{khan_deep_2021}
M Khan, M Gajendran, Y Lee, M Khan, Deep {Neural} {Architectures} for {Medical}
  {Image} {Semantic} {Segmentation}: {Review}.
\newblock {\em\protect\JournalTitle{IEEE Access}} \textbf{9}, 83002--83024
  (2021).

\bibitem{hann_quality_2019}
E Hann, L Biasiolli, Q Zhang, S Neubauer, S Piechnik, Quality control-driven
  image segmentation: Towards reliable automatic image analysis in large-scale
  cardiovascular magnetic resonance aortic cine imaging in {\em Medical {Image}
  {Computing} and {Computer} {Assisted} {Intervention}}.
\newblock pp. 750--758 (2019).

\bibitem{clevert_fast_2016}
DA Clevert, T Unterthiner, S Hochreiter, Fast and {Accurate} {Deep} {Network}
  {Learning} by {Exponential} {Linear} {Units} ({ELUs}) in {\em
  arXiv:1511.07289 [cs]}.
\newblock (2016).

\bibitem{kingma_adam_2017}
D Kingma, J Ba, Adam: {A} {Method} for {Stochastic} {Optimization} in {\em
  arXiv:1412.6980 [cs]}.
\newblock (2017).

\bibitem{chollet_keras_2015}
F Chollet, Keras (2015) https://keras.io.

\bibitem{abadi_tensorflow_2016}
M Abadi, P Barham, J Chen, Z Chen, A Davis, Tensorflow: {A} system for
  large-scale machine learning in {\em Symposium on {Operating} {Systems}
  {Design} and {Implementation}}.
\newblock pp. 265--283 (2016).

\bibitem{byrd_limited_1995}
R Byrd, P Lu, J Nocedal, C Zhu, A {Limited} {Memory} {Algorithm} for {Bound}
  {Constrained} {Optimization}.
\newblock {\em\protect\JournalTitle{Journal on Scientific Computing}}
  \textbf{16}, 1190--1208 (1995).

\end{thebibliography}
\end{document}